\documentclass[aps,prb,twocolumn,superscriptaddress]{revtex4}

\usepackage{graphicx} %I added this.
\begin{document}

%\preprint{}

\title{Superconductivity without Fe or Ni in the phosphides BaIr$_2$P$_2$ and BaRh$_2$P$_2$}


\author{N. Berry}
%\thanks{}
\affiliation{Department of Physics and Astronomy, University of
California Irvine, Irvine CA 92697-4575}

\author{C. Capan}
%\thanks{}
\affiliation{Department of Physics and Astronomy, University of
California Irvine, Irvine CA 92697-4575}

\author{G. Seyfarth}
%\thanks{}
\affiliation{Department of Physics and Astronomy, University of
California Irvine, Irvine CA 92697-4575}\affiliation{D$\acute{e}$partement de Physique, Universit$\acute{e}$ de Montr$\acute{e}$al, Montr$\acute{e}$al H3C 3J7 Canada}

\author{A.D. Bianchi}
%\thanks{}
\affiliation{D$\acute{e}$partement de Physique, Universit$\acute{e}$ de Montr$\acute{e}$al, Montr$\acute{e}$al H3C 3J7 Canada}

\author{J. Ziller}
%\thanks{}
\affiliation{Department of Chemistry, University of California
Irvine, Irvine CA 92697-4575}

\author{Z. Fisk}
%\thanks{}
\affiliation{Department of Physics and Astronomy, University of
California Irvine, Irvine CA 92697-4575}


\date{\today}

\begin{abstract}
Heat capacity, resistivity, and magnetic susceptibility measurements
confirm bulk superconductivity in single crystals of BaIr$_2$P$_2$ (T$_c$=2.1K)
and BaRh$_2$P$_2$ (T$_c$ = 1.0 K). These compounds form in the ThCr$_2$Si$_2$ (122)
structure so they are isostructural to both the Ni and Fe
pnictides but not isoelectronic to either of them. This illustrates
the importance of structure for the occurrence of superconductivity
in the 122 pnictides. Additionally, a comparison between these and
other ternary phosphide superconductors suggests that the
lack of interlayer $P-P$ bonding favors superconductivity.
These stoichiometric and ambient pressure
superconductors offer an ideal playground to investigate the role of
structure for the mechanism of superconductivity in the absence of
magnetism.
\end{abstract}

%\keywords{}

\maketitle

\indent Rare earth intermetallics in the ThCr$_2$Si$_2$(122)
structure have been extensively studied due to their many
interesting properties, such as superconductivity(SC), heavy fermion
behavior, exotic magnetic order, and quantum
criticality\cite{PhysRevLett.43.1892,PhysRevLett.85.626}. The recent
discovery of superconductivity in iron pnictides, first in LaFeAsO
at 26K\cite{Kamihara2008} and soon after in the
AFe$_2$As$_2$(A=Alkali metal) family\cite{rotter:107006}, has
ignited a new interest in non Cu based high $T_{c}$ SC. The ternary
compounds AFe$_2$As$_2$ form in the tetragonal 122 structure and
contain the same building blocks of FeAs planes as LaFeAsO, which
forms in the tetragonal ZrCuSiAs(1111)
structure\cite{rotter:107006,Quebe200070}. Band structure
calculations show a Fermi surface almost exclusively formed by Fe
d-bands\cite{kasinathan-2009}. Fe pnictides are also very tunable
with pressure or chemical substitution, and critical
temperatures($T_{c}$) have reached as high as $55K$ in
SmFeAs(O,F)\cite{0256-307X-25-6-080} and $38K$ in
(Ba,K)Fe$_{2}$As$_{2}$\cite{rotter:107006}. In both families of compounds, SC
is seen to emerge from the suppression of a commensurate
antiferromagnetic order with pressure or
doping\cite{sasmal-2008-101,sefat:117004,0953-8984-21-1-012208}.
Moreover, the long range magnetic order is preceded by (concomitant
to) a structural transition in the 1111 (122)
compounds\cite{kasinathan-2009}. So far, much research has been
focused on the magnetic transition metal elements Fe and Ni with As
in place of Si in the ThCr$_2$Si$_2$ structure. The mechanism for
SC\cite{gordon:127004,nakayama-2008} is still a matter of intensive
debate and investigation in these compounds.

\indent The As atom can be replaced by the isoelectronic element P
forming ternary phosphides in the same 122 structure, as
first investigated by Jeitschko et
al\cite{Jeitschko1987667,Jeitschko1987}. While SC has
not been reported in stoichiometric Fe based ternary phosphides at
ambient pressure, it has been observed in
LaRu$_2$P$_2$\cite{Jeitschko1987}, BaNi$_2$P$_2$\cite{Mine2008}, and
SrNi$_2$P$_2$\cite{Ronnings2009} with $T_{c}$'s $\lesssim$ 4K. Most
ternary phosphides grown with Co exhibit local moment magnetic order
unlike their Fe or Ni counterparts\cite{Reehuis1999}. Isostructural
transitions (tetragonal to collapsed tetragonal) have also been
reported in the ternary phosphides under pressure\cite{Huhnt1997,PhysRevB.56.13796}. Unlike
their As counterparts, these compounds do not show a concomitant magnetic transition\cite{Ronnings2009},
except EuCo$_{2}$P$_{2}$ with its Eu moment ordering\cite{PhysRevLett.80.802}.

\indent This paper reports on single crystal SC in the Co
column for the 122 phosphides, namely in BaIr$_2$P$_2$ and
BaRh$_2$P$_2$. This finding emphasizes the importance of the 122
structure for the stability of SC, since it occurs in the Fe, Co,
and Ni columns of the periodic table. Rh and Ir are non-magnetic
elements in the Co column, between the Fe and Ni columns. This
provides the opportunity to investigate SC without infringing upon
local magnetic moments, known to be detrimental to conventional SC.
Further we show from structural analysis that the interlayer P-P
bonding might be a relevant parameter for the occurrence of SC in
the 122 phosphides.

\indent Single crystals were grown via the standard metal flux
technique\cite{Canfield.1992.PMBOCMSMEOAMP.65}. The single crystals
of BaRh$_2$P$_2$ and CaRh$_2$P$_2$ were grown in Pb flux with a
ratio of 1.3:2:2:40. For BaIr$_2$P$_2$, Cu was added to the Pb flux,
to increase solubility, with molar ratios of
1.3:2:2:40:5(Ba:Ir:P:Pb:Cu). The mixtures were placed inside an
alumina crucible and then sealed in quartz ampoules with inert
atmospheres. All three batches were heated at
$1150^{\circ}C$ for $168h$ and slowly cooled ($4^{\circ}C/h$) to
$450^{\circ}C$, at which point the excess flux was decanted. The
samples were etched in HCl to remove any excess flux. We have also
obtained BaIr$_2$P$_2$ in polycrystalline form from solid state
reaction by mixing stoichiometric amounts of each element and
heating it at $900^{\circ}C$ for $100h$ and then quenched to
300K.

 \begin{figure}
 \includegraphics[width=3in,height=2.75in]{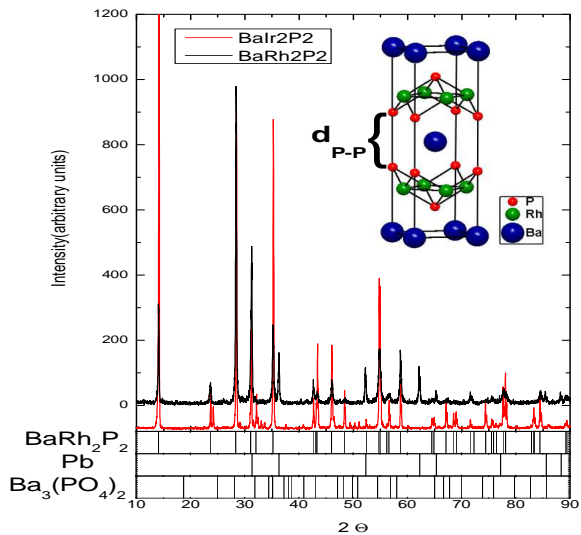}
 \caption{\label{xrays}(Color online)Intensity vs Scattering Angle $\Theta$ obtained
 in powder X-ray diffraction for BaRh$_2$P$_2$(single crystals)
 and BaIr$_2$P$_2$(polycrystals).
 The vertical lines correspond to the reference pattern of BaRh$_2$P$_2$, Pb, and Ba$_{3}$(PO$_{4}$)$_{2}$\cite{Pearson}. The inset represents the tetragonal unit cell of BaRh$_2$P$_2$.}
 \end{figure}

 \begin{table}%[H] add [H] placement to break table across pages
 \caption{Structural Parameters and Physical Properties\label{structparam}}
 \begin{ruledtabular}
 \begin{tabular}{c c c c c c c}
  & a(\r{A}) & c(\r{A}) & $d_{P-P}$(\r{A}) & $T_{c}$(K) & $\gamma(mJ/molK^{2}$) \\
 BaIr$_2$P$_2$ & 3.9469(8) & 12.559(5) & 3.688(2) & 2.1$\pm0.04$ & 9.3$\pm0.6$ \\
 BaRh$_2$P$_2$ & 3.9308(3) & 12.574(2) & 3.725(1) & 1.0$\pm0.04$ & 9.2$\pm0.3$ \\
 CaRh$_2$P$_2$ & 4.0179(3) & 9.655(1) & 2.241(1) & -- & 10.7$\pm0.2$ \\
 \end{tabular}
 \end{ruledtabular}
 \end{table}

\indent The reaction results are first identified by powder X-ray
diffraction. Fig. \ref{xrays} shows the intensity vs scattering
angle $\Theta$ for BaRh$_2$P$_2$ single crystals and for
polycrystalline BaIr$_2$P$_2$ powder. The polycrystalline powder has
a composition of 85\% BaIr$_2$P$_2$, 10\% Ba$_3$(PO$_4$)$_2$ and
only a few percent of Ir$_2$P and IrP$_2$ binaries. Additional peaks
in the BaRh$_2$P$_2$ spectra are from the Pb flux. Single crystals
of both BaRh$_2$P$_2$ and BaIr$_2$P$_2$ are also characterized by a
rotating crystal X-ray diffractometer. The Rietveld refinement
results are shown in table \ref{structparam} and agree with previous
reports\cite{Lohken2002,Wurth1997}. Moreover, the correct
composition and stoichiometry have been confirmed for all single
crystals with Energy Dispersive X-ray Analysis. Magnetic properties
are measured using a commercial SQUID vibrating sample magnetometer.
Heat capacity ($C$) has been measured using a quasiadiabatic heat
pulse technique in a PPMS. The resistivity is measured on a LR700 AC
resistance bridge using Pt wires attached with silver paint. The
single crystals of BaIr$_2$P$_2$ were too small for reliable heat
capacity and magnetization measurements, so these were carried out
on polycrystalline pellets.

 \begin{figure}
 \includegraphics[width=3.4in,height=1.8in]{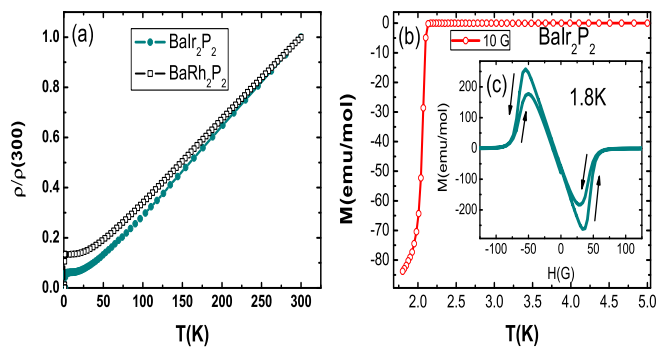}
 \caption{\label{fig_resandsquid}(Color online) (a) Resistivity(normalized) vs Temperature
 in the range $0.5-300K$ in single crystals of BaIr$_2$P$_2$ and BaRh$_2$P$_2$.
 (b) Magnetization vs Temperature in polycrystalline BaIr$_2$P$_2$ showing a diamagnetic jump at $T_c=2.1K$,
 in a field of $10 G$.
 (c) Magnetization vs Magnetic Field(H) at $T=1.8K$ for the same sample.}
 \end{figure}

\indent The temperature ($T$) dependence of resistivity ($\rho$) in
single crystals of BaIr$_2$P$_2$ and BaRh$_2$P$_2$ is shown in fig.
\ref{fig_resandsquid}a from $300K$ down to $0.5K$. The high quality of crystals is evidenced by the large residual resistivity ratios($RRR = \rho(300K)/\rho(3K)$ = 16.5 in BaIr$_2$P$_2$ and 7.5 in BaRh$_2$P$_2$)as well as
by the low values of the residual resistivities ($8.7$ and $1.2\mu\Omega cm$ in Ir and Rh samples). In both systems, $\rho(T)$
exhibits a $T-$linear dependence above $100K$ without any sign of
saturation up to $300K$. There is no evidence for structural or
magnetic transitions in $\rho$ up to $300K$. At low temperatures, a
sharp drop to $\rho=0$ indicates the onset of SC at $T_c=2.1K$ in
BaIr$_2$P$_2$ and $1K$ in BaRh$_2$P$_2$.

\indent The temperature and magnetic field dependence of
magnetization (M) are shown in fig. \ref{fig_resandsquid}b,c for
polycrystalline BaIr$_2$P$_2$. The diamagnetic jump in M(T)
corresponds to the same $T_c$ as determined from $\rho(T)$. The
magnetization loop M(H) at $1.8K$ shows hysteresis and rather broad
extrema. The average of their field positions (necessary due to
trapped flux in the magnet) yields 45 Oe as an upper bound for the
lower critical field $H_{c1}$ (we adopt the type-II SC scheme since
the critical field determined by $\rho(T,H)$ and $C(T,H)$ is
significantly higher, see below). The slope of M(H) below $H_{c1}$
is used to estimate that 100\% of the volume is superconducting.

\begin{figure}
\includegraphics[width=3.5in,height=3in]{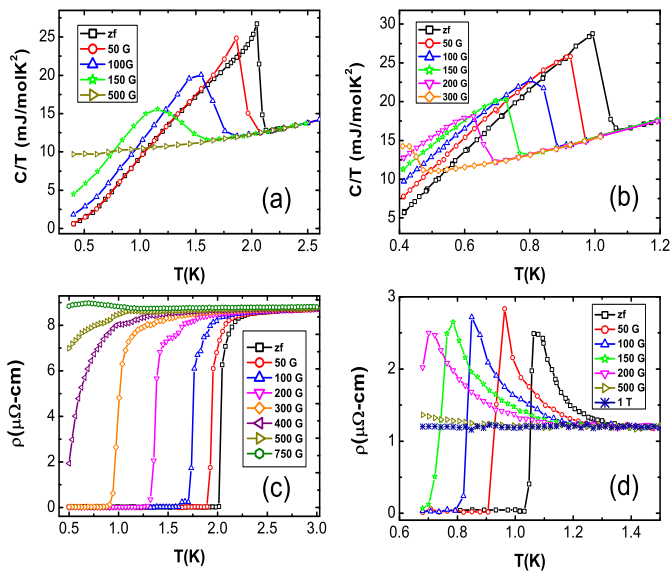}
\caption{\label{fig_fielddep}(Color online) Total $\frac{C}{T}(T)$ at the indicated magnetic fields in polycrystalline
BaIr$_2$P$_2$(a) and single crystals of BaRh$_2$P$_2$(b). Resistivity vs Temperature at the
indicated magnetic fields in single crystals of BaIr$_2$P$_2$(c)
and BaRh$_2$P$_2$(d).}
\end{figure}

\indent The bulk nature of SC is also confirmed with a sharp anomaly
in $C(T)$, observed in both compounds (fig. \ref{fig_fielddep}a,b).
The good agreement between the thermodynamic and resistive $T_c$ and
the sharpness of the transition even for the polycrystalline sample
imply that $T_c$ does not show any distribution. At
zero field, the ratio $\frac{\Delta C}{\gamma T_c}$ equals 1.41 and
1.17 for BaIr$_2$P$_2$ and BaRh$_2$P$_2$, consistent with BCS theory. The values of the electronic specific heat
coefficient $\gamma$ are obtained from a linear fit to $\frac{C}{T}$
vs $T^2$ in the range 0.4-2.4K(BaIr$_{2}$P$_{2}$) and 0.5-1.0K(BaRh$_{2}$P$_{2}$)(see table \ref{structparam}). The $\gamma$ and T$_{c}$ 
shown here are consistent with those reported on polycrystals\cite{JPSJ.78.023706}. 

\begin{figure}
\includegraphics[width=2in,height=2in]{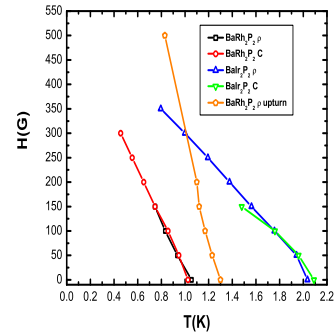}
\caption{\label{fig_hc2}(Color online) Upper critical field ($H_{c2}$) vs
Temperature in BaIr$_2$P$_2$ and in BaRh$_2$P$_2$, as determined
from resistivity and specific heat data. Also shown is the onset of
the resistivity upturn in BaRh$_2$P$_2$.}
\end{figure}

\indent The suppression of $T_{c}$ with magnetic field is seen in
$\frac{C}{T}(T)$ (fig. \ref{fig_fielddep}a,b) and $\rho(T)$
 (fig. \ref{fig_fielddep}c,d). In both compounds, the superconducting
transition in $\rho(T)$ remains rather sharp, even under magnetic
fields as high as $200 Oe\approx\frac{H_{c2}}{2}$. This suggests a
rather strongly pinned vortex lattice. The specific heat anomaly
also remains sharp for the BaRh$_2$P$_2$ single crystals up to $150
Oe$ (see fig. \ref{fig_fielddep}b), but broadens with field for the
polycrystalline BaIr$_2$P$_2$ (see fig. \ref{fig_fielddep}a). The
possible anisotropy of the upper critical field has not been
investigated and might be responsible for this broadening. The
corresponding $H-T$ phase diagram is shown in fig. \ref{fig_hc2}.
There is a good agreement between the values obtained from
resistivity and specific heat for both compounds. The use of the
approximation $H_{c2}(0)\simeq-0.7T_{c}\frac{\partial
H_{c2}}{\partial T}|_{Tc}$ yields $H_{c2}(0)=410 Oe$ and $370 Oe$ in
BaIr$_2$P$_2$ and BaRh$_2$P$_2$. These values of $H_{c2}(0)$ are
comparable to SrNi$_2$P$_2$($390 Oe$)\cite{Ronnings2009} and 
BaNi$_2$P$_2$($550 Oe$)\cite{Mine2008},
but smaller than those cited in Hirai et al\cite{JPSJ.78.023706}. The broadness
of the transitions in polycrystals\cite{JPSJ.78.023706} may be the source
of the discrepancies. 
From our values of $H_{c2}(0)$ we estimate the coherence
lengths to be 80nm and 95nm for BaIr$_2$P$_2$ and BaRh$_2$P$_2$
respectively.

\indent Fig. \ref{fig_fielddep}d shows a pronounced upturn in
$\rho(T)$ preceding the onset of the superconducting jump in
BaRh$_2$P$_2$. The resistivity rises about 100$\%$ in the
temperature interval 1.35-1K at zero field. The onset of the upturn
is suppressed with magnetic field but its amplitude is unaffected.
Moreover, this suppression does not appear to be correlated with the
upper critical field $H_{c2}(T)$, as seen in fig. \ref{fig_hc2},
suggesting separate phenomena. We have verified that the upturn is
present in a second crystal of BaRh$_2$P$_2$ of similar RRR, as well
as in a polycrystalline pellet, but found that the amplitude of the
upturn is sample-dependent. A smaller upturn is also observed in
single crystal BaIr$_2$P$_2$ above $700 Oe$, but it is absent at
zero field in this case. Such an upturn is also reported in
SrNi$_2$P$_2$\cite{Ronnings2009} and LaFePO\cite{analytis-2008}.  In
addition, a sample dependent Curie tail has been frequently observed
in the low temperature susceptibility, with an associated Brillouin
like behavior in M vs H for single crystals of both compounds (not
shown). The sample-to-sample variation of this magnetic behavior is
suggestive of an extrinsic origin, although the corresponding
concentration of spin 1/2 is far in excess of the level of magnetic
impurities contained in the starting materials ($\leq 20ppm$). Their
origin remains unclear and is beyond the scope of this paper.

\indent We now turn to the relationship between the 122
structure and SC in pnictides. BaFe$_2$As$_2$
is a prime example of the flexibility of this structure on the route
to SC: it has been shown that pressure and doping on
all three atomic sites have independently induced SC\cite{rotter:107006,sefat:117004,Jiang2009,0953-8984-21-1-012208}.
However, in the isoelectronic CaFe$_2$As$_2$ the existence of SC is highly
controversial\cite{kreyssig:184517,yu:020511}.
Moreover, in CaFe$_2$As$_2$, recent theoretical calculations
show an intimate connection between the Fe-spin state
and the interlayer $As-As$ bonding\cite{yildirim:037003}. Future studies should clarify
the effect of the Fe moments on $T_c$. 
In addition, BaRh$_{2}$As$_{2}$ exhibits no superconductivity down to 1.8K\cite{singh:104512}.
Thus, the relationship between the
tetragonal structure and SC is not clear
at present in the 122 arsenides. 
The present phosphides allow the investigation of the relationship between SC and structure
without the interference of magnetism, since neither Rh nor Ir are intrinsically magnetic.

\indent In the ternary phosphides there is an isostructural
transition into a collapsed tetragonal structure\cite{Huhnt1997}, similar to
CaFe$_{2}$As$_{2}$\cite{kreyssig:184517,yu:020511},
except that it does not appear to be associated with any magnetic
order\cite{Ronnings2009}. Previous investigations in BaRh$_2$P$_2$
did not show any structural transition up to $11GPa$ and down to low
temperatures\cite{Huhnt1997}(no reports for BaIr$_2$P$_2$). In the phoshpides,
unlike the Arsenides, proximity to a structural transition is not a
pre-requisite for SC. It is known that this
isostructural transition corresponds to the formation or breaking of
a bond between the interlayer P atoms\cite{Banu2000}. In the absence
of $P-P$ bond, the cohesion of the layers is due to the Coulomb
attraction through the intermediate $A^{2+}$
cation\cite{Hoffman1985}. The critical distance for bond formation
obtained theoretically is about $d_{c} \sim 2.8{\AA}$ between the
interlayer P atoms, labeled $d_{P-P}$ in the inset of fig.
\ref{xrays}\cite{Hoffman1985}. We found that both BaRh$_2$P$_2$ and
BaIr$_2$P$_2$ have a $d_{P-P}$ of $\sim3.7{\AA}$ (see table
\ref{structparam}) indicating the absence of interlayer bonding
between the P atoms, which is consistent with structural
calculations\cite{Banu2000}. In contrast, CaRh$_2$P$_2$ has a
$d_{P-P}$ of only $2.25{\AA}$, which is below the critical distance
for bond formation. We have also grown single crystals of
CaRh$_2$P$_2$ and found no evidence of SC down to $0.55K$. This
suggests that the absence of bonds favors SC.

\indent The absence of $P-P$ bonds is also found in other
superconducting phosphides, such as BaNi$_2$P$_2$ which has
$d_{P-P}=3.71{\AA}$\cite{Keimes2004}. In fact, none of the known
ternary phosphides (BaIr$_2$P$_2$, BaRh$_2$P$_2$,
BaNi$_2$P$_2$\cite{Mine2008}, and LaRu$_2$P$_2$\cite{Jeitschko1987})
that exhibit ambient pressure SC in the tetragonal structure are
bonded between the interlayer P atoms. Nevertheless, it is
interesting that LaRu$_2$P$_2$, with the highest $T_c$ of $4.1K$,
lies closest ($3.00{\AA}$) to the theoretical structural
instability, while still being in the non-bonding
state\cite{Jeitschko1987}. However, SrNi$_2$P$_2$ shows SC in the
collapsed tetragonal phase under pressure where a bond exists
between the layers\cite{Ronnings2009}. Since the ambient pressure
orthorhombic phase is also superconducting it is hard to assess the
importance of the structure for SC in this case. De Haas-van Alphen
results of BaNi$_2$P$_2$ show a 3D Fermi surface dominated by the Ni
d-bands, indicating that the effect of interlayer coupling on the
electronic dimensionality is small\cite{JPSJ.78.033706}. Our results
lay the groundwork for more theoretical investigations in order to clarify
the relationship between the interlayer bonding and the mechanism
for SC in the non-magnetic 122 phosphides.

In conclusion, we have shown the existence of bulk weak coupling
SC in 122 pnictides in the Co column of the periodic table with
non-magnetic transition metals Rh and Ir. This emphasizes the
importance of the 122 structure and the robustness of
SC with respect to changes in the electronic
configuration, opening the door for SC
in other non Fe based compounds. Also, these findings suggest that the lack of interlayer bonding
favors SC. It is important to
understand how the structure affects SC in the
ternary and quaternary pnictides in the
absence of competing magnetic order. Due to the apparent lack of
magnetism, BaIr$_2$P$_2$ and BaRh$_2$P$_2$ provide
convenient systems in which to study the interplay between structure
and SC.

%\appendix
%\section{}

\begin{acknowledgments}
This research was supported with funding from NSF DMR 0854781.
A.D.B. received support from Natural Sciences and Engineering
Research Council of Canada, Fonds Qu$\acute{e}$b$\acute{e}$cois de
la Recherche sur la Nature et les Technologies, and Canada Research
Chair Foundation.
\end{acknowledgments}

%\bibliography{paper-nourl}

\end{document}